\documentclass[prl,floatfix,twocolumn]{revtex4}
\usepackage{bm}
\usepackage{graphicx}
\usepackage{amsmath,amssymb,amsgen,amsbsy}
\usepackage{color}
\usepackage{epstopdf}

\allowdisplaybreaks

\begin{document}

\title{Emergence of chaos in a viscous solution of rods}
% for header on odd pages
\author{Emmanuel L. C. VI M. Plan}
%\affiliation{Universit\'e C\^ote d'Azur, CNRS, LJAD, Nice, France}
\author{Stefano Musacchio}
%\affiliation{Universit\'e C\^ote d'Azur, CNRS, LJAD, Nice, France}
\author{Dario Vincenzi}
\affiliation{Universit\'e C\^ote d'Azur, CNRS, LJAD, Nice, France}

\date{\today}

\begin{abstract}
It is shown that the addition of small amounts of
microscopic rods in a viscous fluid at low Reynolds number causes
a significant increase of the flow resistance. Numerical simulations
of the dynamics of the solution reveal that this phenomenon is associated
to a transition from laminar to chaotic flow. Polymer stresses give rise
to flow instabilities which, in turn, perturb the alignment of the
rods. This coupled dynamics results in the activation of a wide
range of scales, which enhances the mixing efficiency of viscous flows.
\end{abstract}
 
%%%%%%%%%%%%%%%%%%%%%%%
%\pacs{...}

\maketitle
%The dynamics of viscous fluids at low Reynolds number is attracting an increasing 
%interest in a broad variety of applications 
%ranging from the biophysics of planktonic microorganisms\cite{GRS12} 
%to the dynamics of Fermi-liquids in graphene monolayers\cite{LF16}. 
 
In a laminar flow the dispersion of substances occurs by molecular diffusion, 
which operates on extremely long time scales.
Various strategies have therefore been developed, particularly in microfluidic
applications, to accelerate mixing
and dispersion at low fluid inertia \cite{SQ05,OW04,SSA04}.
The available strategies are commonly
divided into two classes, passive or active, according to
whether the desired effect is obtained through the specific
geometry of the flow 
or through an oscillatory forcing within the fluid \cite{SSA04}.
An alternative method for improving the mixing properties of
low-Reynolds-number flows was proposed by Groisman and Steinberg \cite{GS01}
and consists in adding elastic polymers to the fluid.
If the inertia of the fluid is low but 
the elasticity of polymers is large enough, elastic
stresses give rise to instabilities that ultimately
generate a chaotic regime
known as ``elastic turbulence'' \cite{GS00}.
In this regime the velocity field, although remaining smooth in space, 
becomes chaotic and develops a power-law energy spectrum, 
which enhances the mixing properties of the flow.
While the use of elastic turbulence in microfluidics is now well established \cite{BSBGS04,JV11,PBCS12,TCT15,AWDP16},
new potential applications have recently emerged, namely
in oil extraction from porous rocks \cite{Clarke16}.

In this Letter we propose a novel mechanism for generating chaotic flows at low Reynolds 
numbers that does not rely on elasticity. It is based on the addition of rigid rodlike polymers. 
At high Reynolds numbers, 
elastic- and rigid-polymer solutions exhibit remarkably similar macroscopic behavior
(\textit{e.g.}, Refs. \cite{VSW97,PDDS04,BCLLP05, G08}).
In both cases the turbulent drag is considerably reduced compared to that of the solvent alone. 
% according to a phenomenology that seems to depend little on the microscopic structure of the solution. 
In particular, when either type of polymer is added in sufficiently high concentrations to a turbulent 
channel flow of a Newtonian fluid,
the velocity profile continues to depend logarithmically on the distance from the walls of the channel, but the mean velocity increases to a value known as maximum-drag-reduction asymptote.
%as a function both of the Reynolds number and of polymer concentration 
%until an asymptotic mean profile, known as maximum-drag-reduction asymptote, is reached.
Here we study whether or not the similarity between elastic- and rigid-polymer solutions 
carries over to the low-Reynolds-number regime, 
\textit{i.e} whether or not the addition of rigid polymers originates   
a regime similar to elastic turbulence. 

We consider a dilute solution of inertialess rodlike polymers.
The polymer phase is described by the 
symmetric unit-trace tensor field $\mathcal{R}(\bm x,t)=\overline{n_i n_j}$, 
where $\bm n$ is the orientation of an 
individual polymer and the average is taken over the polymers contained in a volume element at position $\bm x$ at time $t$.  
The coupled evolution of $\mathcal{R}(\bm x,t)$ and the incompressible velocity field $\bm u(\bm x,t)$ is given by the following equations \cite{DE88,PLB08} (summation over repeated indices is implied):
\begin{subequations}
\begin{alignat}{4}
\partial_t {u_i} + {u_k}\partial_k{u_i} &= -\partial_i p + \nu \partial^2{u_i} + 
\partial_k{\sigma_{ik}} + {f_i}\,,\label{09_eq:sys1a}
\\
\partial_t{\mathcal{R}_{ij}} + {u_j} \partial_j{\mathcal{R}_{ij}} &= (\partial_k{u_i}){\mathcal{R}_{kj}}
 + {\mathcal{R}_{ik}}(\partial_k{u_j})- \nonumber\\
&\hspace{1cm} 2\mathcal{R}_{ij}(\partial_l u_k) \mathcal{R}_{kl},
\label{09_eq:sys1b}
\end{alignat}
\label{09_eq:sys1}
\end{subequations}
where $\partial_k=\partial/\partial x_k$, $p(\bm x,t)$ is pressure, 
$\nu$ is the kinematic viscosity of the fluid, and $\bm f(\bm x,t)$ 
is the body-force which sustains the flow. 
The polymer stress tensor takes the form 
%\begin{equation}
$\sigma_{ij}=6\nu \eta_\mathrm{p}\mathcal{R}_{ij}(\partial_l u_k)\mathcal{R}_{kl}$ \cite{DE88}. 
%\label{09_eq:stresstensor}
%\end{equation}
The intensity of the polymer feedback on the flow is determined by the polymer concentration,
which is proportional to $\eta_\mathrm{p}$. 
This expression for the polymer stress tensor is based 
on a quadratic approximation proposed by Doi and Edwards \cite{DE88}. 
More sophisticated closures have been employed in the literature (see,
\textit{e.g.}, Ref. \cite{MHJS11} and references therein); 
here we focus on the simplest model of rodlike-polymer solution
that may display instabilities at low Reynolds number. 
In addition, we disregard Brownian rotations
assuming that the orientation of polymers
is mainly determined by the velocity gradients.    

For large values of the Reynolds number, the system described by Eqs. \eqref{09_eq:sys1}
has been shown to reproduce the main features of drag reduction in turbulent solutions of rodlike polymers \cite{BCLLP05,BCDP08,PLB08,ABKLLP08}.
Here we study the same system at small values of the Reynolds number. 
Equations \eqref{09_eq:sys1} are solved over a two-dimensional $2\pi$-periodic box 
and $\bm f$ is taken to be the Kolmogorov force 
$\bm f(\bm x)=(0,F\sin(x/L))$.
For $\eta_\mathrm{p}=0$ the flow has the laminar solution $\bm u=(0,U_0 \sin(x/L))$ 
with $U_0=FL^2/\nu$, which becomes unstable when the Reynolds number
$\mathrm{Re}=U_0L/\nu$ exceeds the critical value $\mathrm{Re}_\mathrm{c}=\sqrt{2}$ and 
eventually turbulent when $\mathrm{Re}$ is increased further
(\textit{e.g.}, Ref. \cite{MB14}). 
Even in the turbulent regime, the mean flow has the sinusoidal form $\langle\bm u\rangle=(0,U \sin(x/L))$, where $\langle\bm\cdot\rangle$  denotes an average over the variable $y$ and over time.
The Kolmogorov force has been previously used 
in the context of non-Newtonian fluid mechanics
to study turbulent drag reduction \cite{BCM05}, the formation of low-$\mathrm{Re}$ instabilities in viscoelastic \cite{BCMPV05,BBCMPV07} and rheopectic fluids \cite{BMP13}, and elastic turbulence \cite{BBBCM08,BB10}.

Numerical simulations of Eqs.~\eqref{09_eq:sys1} 
are performed by using a dealiased pseudospectral method with $1024^2$ gridpoints.
The time-integration uses a fourth-order Runge-Kutta scheme with
implicit integration of the linear dissipative terms. 
The parameters of the simulations are set to 
keep $\mathrm{Re}=1$ fixed below $\mathrm{Re}_\mathrm{c}$ 
in the absence of polymer feedback ($\eta_\mathrm{p}=0$). 
The viscosity is set to $\nu=1$, 
the length scale of the forcing is either $L=1/4$ or $L=1/8$, 
and its amplitude is $F=\nu^2/L^3$. 
The feedback coefficient is varied from $\eta_{\mathrm{p}}=1$ to $\eta_{\mathrm{p}}=5$. 
The stiffness of the equations increases with $\eta_\mathrm{p}$, 
limiting the accessible range of parameters. 

\begin{figure}
\centering\includegraphics[width=0.23\textwidth]{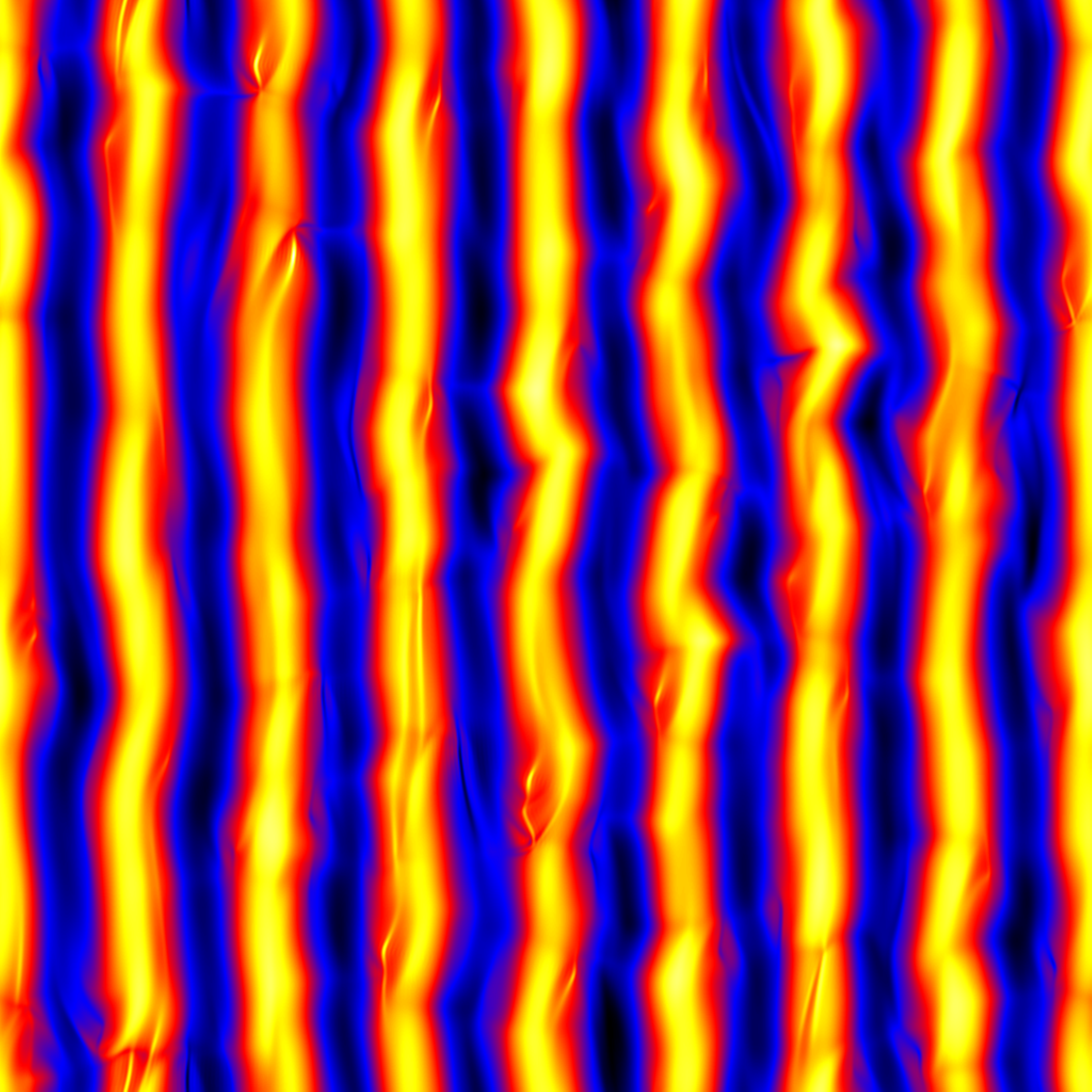}\hfill%
\includegraphics[width=0.23\textwidth]{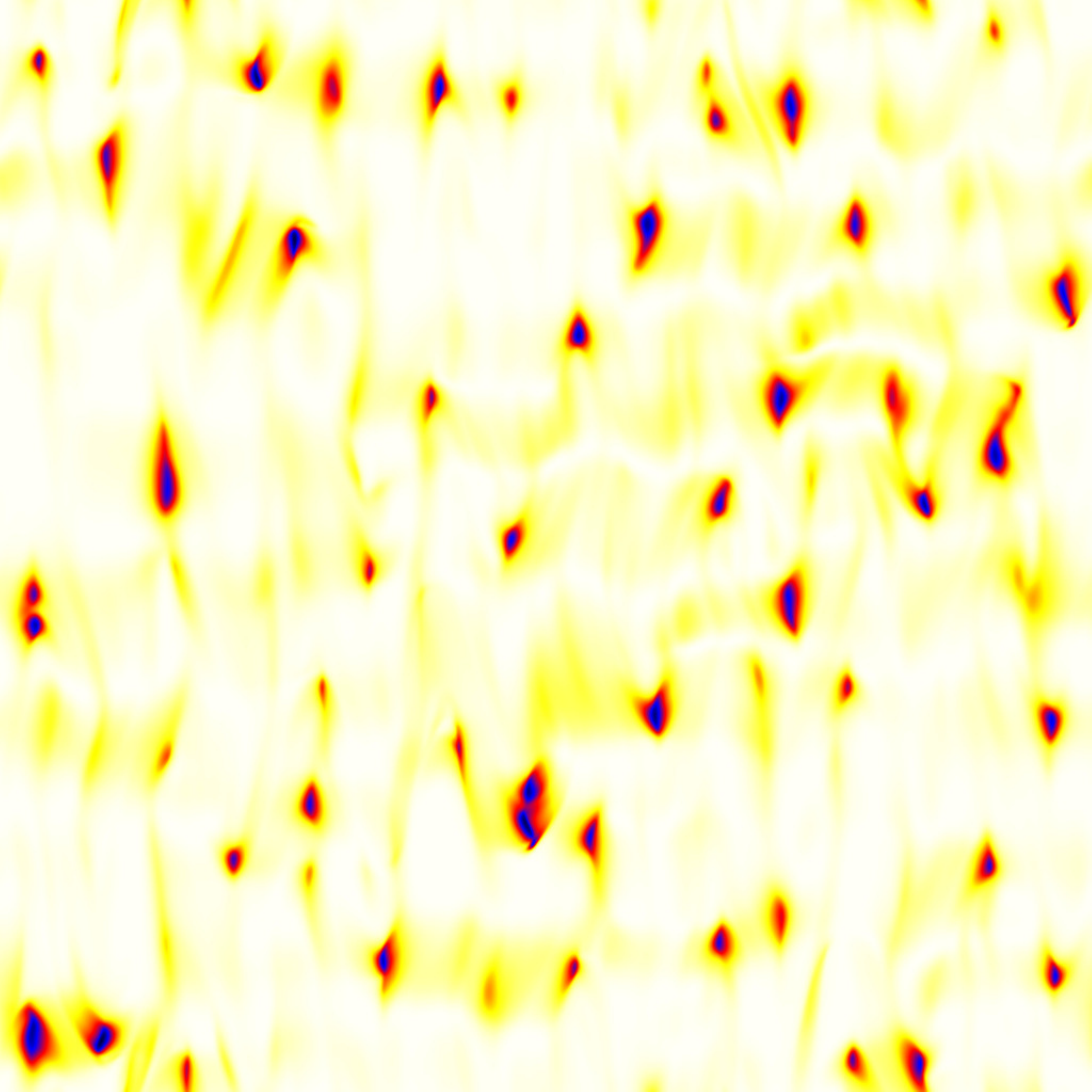}
\caption{Left: Snapshot of the vorticity field $\omega$ for 
  $\eta_\mathrm{p}=3$ and $L=1/8$. Black (white) represents negative
  (positive) vorticity.
Right: Snapshot of the component $\mathcal{R}_{11}$. White represents 0, black represents~1.
}
\label{09_fig1}
\end{figure}

\begin{figure}
\centering\includegraphics[width=0.47\textwidth]{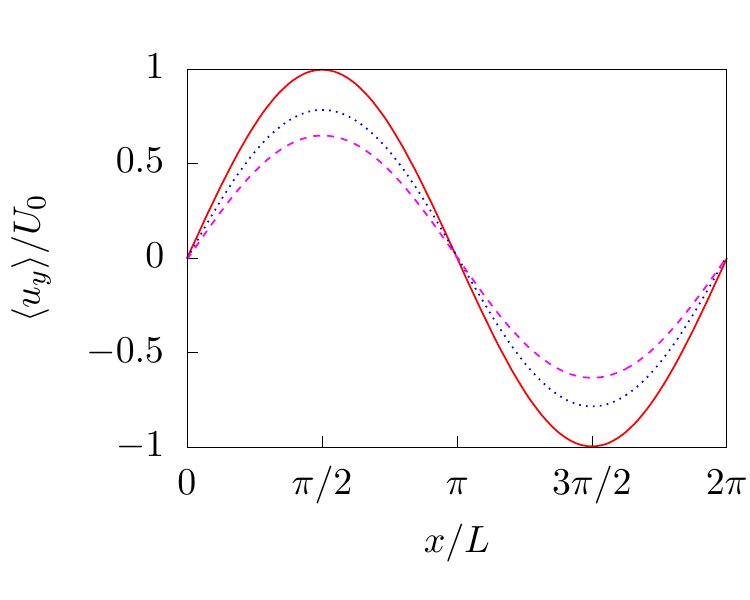}
\caption{Mean velocity profiles for $L=1/8$ and concentrations
$\eta_{\mathrm{p}}=1$ (solid), 
$\eta_{\mathrm{p}}=3$ (dotted), and 
$\eta_{\mathrm{p}}=5$ (dashed).}
\label{09_fig1b}
\end{figure}

Initially the flow is a weak perturbation of the $\eta_\mathrm{p}=0$ stable solution, 
while the components of $\mathcal{R}$ are randomly distributed.
%Different choiches of the initial condition for the polymer phase have been investigated, 
%leading to similar results after an initial transient. 
When the feedback of the polymers is absent ($\eta_\mathrm{p}=0$) 
the initial perturbation decays
and the polymers align with the direction of the mean shear flow.
Conversely, at large $\eta_\mathrm{p}$ the flow is 
strongly modified by the presence of the rods. 
The streamlines wiggle over time and thin filaments appear in the 
vorticity field $\omega=|\bm\nabla\times\bm u|$ (see
Fig. \ref{09_fig1}, left panel).
These filaments correspond to appreciable localized perturbations of the tensor $\mathcal{R}$ 
away from the laminar fixed point (Fig. \ref{09_fig1}, right panel) 
and are due to the rods being unaligned with the shear direction.
Notably, we find that the mean flow, obtained by means of long time averages, 
maintains the sinusoidal form $\langle\bm u\rangle=(0,U \sin(x/L))$ 
also in the presence of strong polymer feedback (Fig. \ref{09_fig1b}). 

\begin{figure}
\centering\includegraphics[width=0.47\textwidth]{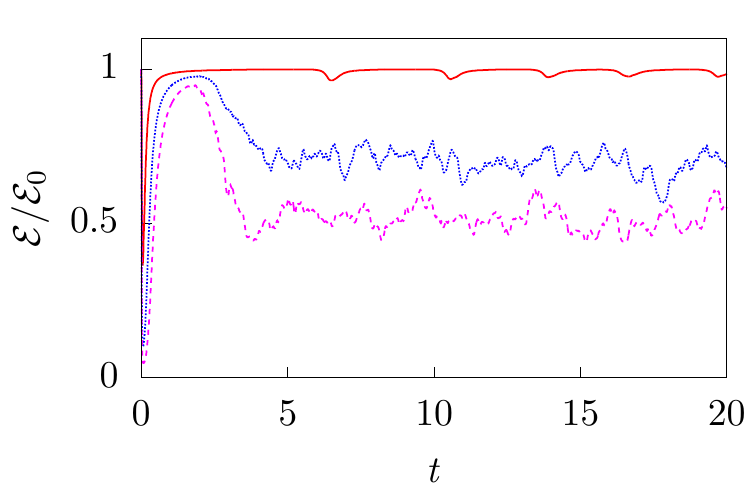}
\caption{The kinetic energy $\mathcal{E}$ for $L=1/8$ and
concentrations $\eta_{\mathrm{p}}=1$ (top, solid), $\eta_{\mathrm{p}}=3$ (middle, dotted), and $\eta_{\mathrm{p}}=5$ (bottom, dashed) divided by the kinetic energy $\mathcal{E}_0=F^2L^4/2\nu^2$ corresponding to $\eta_\mathrm{p}=0$ and the same value of the force $F$.}
\label{09_fig2}
\end{figure}

The time series of the kinetic energy in Fig. \ref{09_fig2} show that, in the case of a low concentration ($\eta_{\mathrm{p}}=1$), the system repetitively attempts but fails to escape the laminar regime in a quasiperiodic manner. The amount of kinetic energy is initially close to that in the laminar regime. After some time, the solution dissipates a small fraction of kinetic energy but quickly relaxes back towards the laminar regime until it restarts this cyclic pattern.
In contrast, for higher concentrations the kinetic energy is significantly reduced and, after an initial transient, fluctuates around a constant value. We have observed that different initial conditions for $\mathcal{R}$ may give rise to longer transients that involve a quasiperiodic sequence of activations and relaxations comparable to that observed for low values of $\eta_{\mathrm{p}}$. Nevertheless, the statistically steady state achieved at later times is independent of the peculiar choice of initial conditions. 

\begin{figure}
\centering\includegraphics[width=0.39\textwidth]{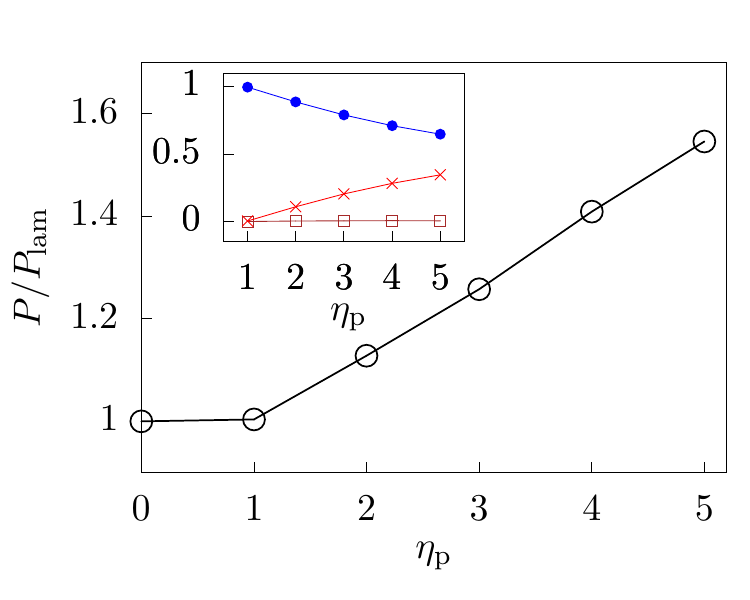}
\caption{The normalized mean injected power $P/P_\mathrm{lam}$ as a function of $\eta_{\mathrm{p}}$. 
Inset: The amplitudes of the stresses $\Pi_r$ (brown $\square$), $\Pi_\nu$ (blue ${\bullet}$), and $\Pi_p$ (red $\times$) divided by the amplitude of the total stress $\Pi_\mathrm{tot}$ for $L=1/8$ and different values of $\eta_{\mathrm{p}}$ (see Table \ref{09_tab:stress}).}
\label{09_fig3}
\end{figure}

The reduction of the kinetic energy of the flow at fixed intensity of the external force 
reveals that the presence of the rods causes an increase in the flow resistance. 
This effect can be quantified by the ratio of the actual mean power $P=FU/2$ 
provided by the external force and the power $P_\mathrm{lam}=F_0U/2$ that would be required to 
sustain a laminar mean flow with the same amplitude $U$ in the
absence of polymers. 
In the latter case, the force required would be $F_0=\nu U/L^2$ and 
the corresponding mean power would be $P_\mathrm{lam}=F_0U/2 = \nu U^2/2 L^2$.
Figure \ref{09_fig3} shows the ratio
\begin{equation}
\dfrac{P}{P_\mathrm{lam}}=\dfrac{F}{F_0}=\dfrac{FL^2}{\nu U}
\end{equation}
as a function of $\eta_\mathrm{p}$ and indicates that more power is required 
to sustain the same mean flow in solutions with higher concentrations.

The analysis of the momentum budget confirms 
that the increased resistance is due to an increase of the 
amount of stress due to the polymers. 
In the steady state the momentum budget 
can be obtained by averaging Eq.~\eqref{09_eq:sys1a} over $y$ and time:
\begin{equation}
\partial_x \Pi_r=\partial_x(\Pi_\nu+\Pi_p)+f_y,
\label{09_eq:budget}
\end{equation}
where $\Pi_r=\langle u_xu_y \rangle$, $\Pi_\nu=\nu\partial_x\langle u_y\rangle$, and $\Pi_p=\langle \sigma_{xy}\rangle$ are the Reynolds, viscous, and polymer stress, respectively. 
Remarkably we find that these profiles remain sinusoidal as in the $\eta_\mathrm{p}=0$ case, 
namely $\Pi_\mathrm{r}=-S\cos(x/L)$, $\Pi_\nu=\nu UL^{-1}\cos(x/L)$, 
and $\Pi_p=\Sigma\cos(x/L)$. 
Equation \eqref{09_eq:budget} then yields the following relation between 
the amplitudes of the different contributions to the stress:
\begin{equation}
S+\dfrac{\nu U}{L}+\Sigma=FL.
\label{09_eq:coefficients}
\end{equation}
These contributions are reported in Table \ref{09_tab:stress} 
and they are shown in the inset of Fig. \ref{09_fig3}.
\begin{table}
\begin{tabular}{|c|c|c|c|c|}\hline
 $\eta_\mathrm{p}$ & $L$ & $\Pi_r/\Pi_\mathrm{tot}$ & $\Pi_\nu/\Pi_\mathrm{tot}$ & $\Pi_p/\Pi_\mathrm{tot}$ \\ \hline
1 & 1/8 & 0.001 & 0.996 & 0.004 \\
2 & 1/8 & 0.004 & 0.887 & 0.110 \\
3 & 1/4 & 0.005 & 0.787 & 0.209 \\
3 & 1/8 & 0.005 & 0.795 & 0.200 \\
4 & 1/8 & 0.007 & 0.710 & 0.284 \\
5 & 1/8 & 0.006 & 0.647 & 0.347 \\ \hline
\end{tabular}
\vspace{2mm}
\caption{Amplitude of the stresses $\Pi_r$, $\Pi_\nu$, and $\Pi_p$ 
divided by the amplitude of the total stress $\Pi_\mathrm{tot} = \Pi_r + \Pi_\nu + \Pi_p$ 
for different values of $\eta_\mathrm{p}$ and $L$.}
\label{09_tab:stress}
\end{table}
The results confirm that the polymer contribution to the total stress increases with $\eta_\mathrm{p}$, 
whereas that of the viscous stress decreases.
The contribution of the Reynolds stress is extremely small (less than $10^{-2}$), 
which demonstrates that inertial effects remain negligible as $\eta_\mathrm{p}$ is increased.
Figures \ref{09_fig2} and \ref{09_fig3} also suggest 
the presence of a threshold concentration for the appearance of 
fluctuations. 

Further insight into the dynamics of the solution is gained by examining the energy balance in wave-number space.
\begin{figure}
\centering\includegraphics[width=0.47\textwidth]{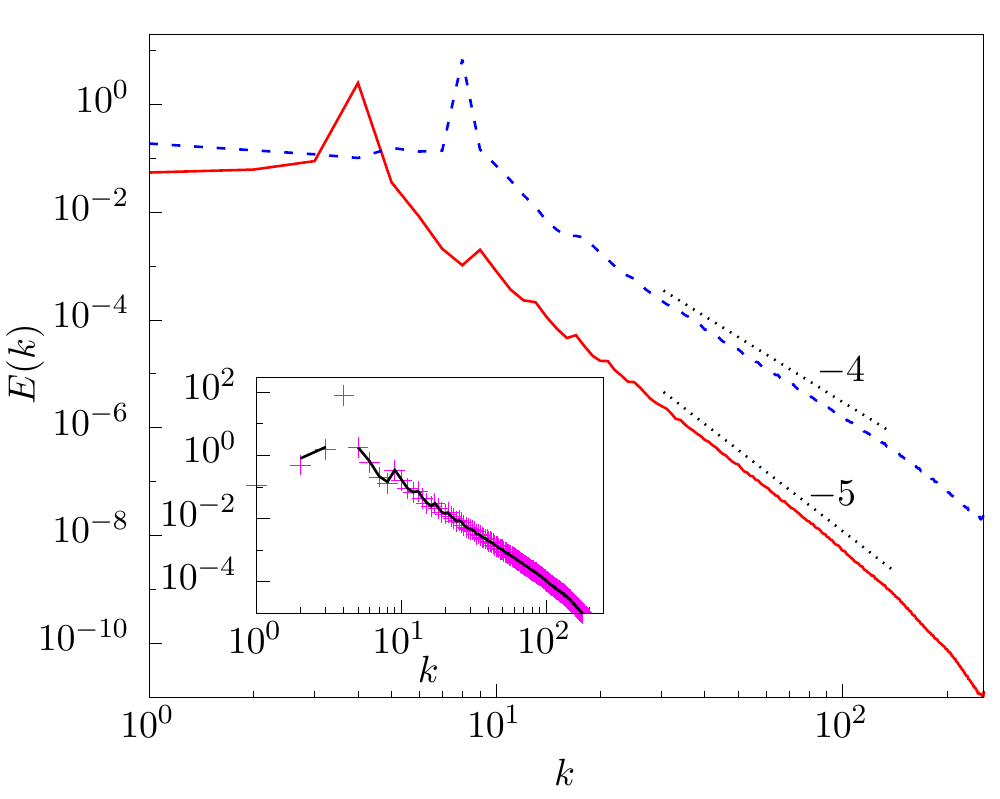}
\caption{The kinetic energy spectrum $E(k)$ for $L=1/4$, $\eta_{\mathrm{p}}=3$ (solid red) and $L=1/8$, $\eta_{\mathrm{p}}=5$ (dashed blue). The two black dotted segments represent $k^{-4}$ and $k^{-5}$. Inset: The unsigned kinetic-energy dissipation spectrum $2\nu k^2E(k)$ (magenta $+$) and the polymer energy transfer (solid black) for $L=1/4$ and $\eta_{\mathrm{p}}=3$.}
\label{09_fig5}
\end{figure}
For sufficiently large values of $\eta_{\mathrm{p}}$, the kinetic-energy spectrum 
behaves as a power law $E(k)\sim k^{-\alpha}$, where the exponent $\alpha$ depends both 
on the concentration and on the scale of the force and varies between $4$ and $5$ (Fig. \ref{09_fig5}). 
A wide range of scales is therefore activated, 
and this results to an enhancement of the mixing properties of the flow.
Furthermore, the energy transfer due to the fluid inertia is negligible, 
and the dynamics is characterized by a scale-by-scale balance between the polymer 
energy transfer and viscous dissipation (inset of Fig. \ref{09_fig5}). 

The regime described here has properties comparable to those of elastic turbulence in viscoelastic fluids, 
namely the flow resistance is increased with the addition of rods and 
the kinetic-energy spectrum displays a power-law steeper than $k^{-3}$.
In addition the Reynolds stress and the energy transfer due to the fluid inertia
are negligible; hence the emergence of chaos is entirely attributable to polymer stresses.
Our study establishes an analogy between the behavior of viscoelastic 
fluids and that of solutions of rodlike polymers, 
similar to what is observed at high Reynolds number. 
%It is worth to notice that the system considered here cannot be obtained trivially from 
%viscoelastic models in the limit of vanishing relaxation time.
These results therefore demonstrate that elasticity is not essential 
to generate a chaotic behavior at low Reynolds numbers and
indicate an alternative mechanism to enhance mixing in microfluidic flows. 
This mechanism presumably has the advantage of being less affected by the 
degradation observed in elastic turbulence \cite{GS04}, 
since there are experimental evidences that the degradation due to large strains is weaker for rodlike polymers
than for elastic polymers \cite{PAS13}.

Experimental studies aimed at investigating the phenomenon proposed in this Letter would be very interesting. 
Open questions concern the dependence of the mixing properties of rigid-polymer solutions on the type of force and on the boundary conditions. 
Additional insight into the dynamics of these polymeric fluids
would also come from a stability analysis of system \eqref{09_eq:sys1}, in the spirit of the approach
taken for the study of low-Reynolds-number instabilities in viscoelastic \cite{BCMPV05,BBCMPV07} and rheopectic \cite{BMP13} fluids.
Finally, the orientation and rotation statistics of microscopic rods in turbulent flows has recently attracted a lot of attention \cite{PW11,PCTV12,GEM14,GVP14,VS17}; it would be interesting to investigate the dynamics of individual rods in the flow regime studied here.

\begin{acknowledgments}
The authors would like to acknowledge the support of the EU COST Action MP 1305 `Flowing Matter.' The work of E.L.C.M.P. was supported by EACEA through the Erasmus
Mundus Mobility with Asia program.
\end{acknowledgments}
%
%\oneappendix
%\section{Derivation of the decay rate of $N$-dimensional volumes}
%\label{appA}

%%%%%%%%%%%%%%%%%%%%%%

\end{document}